\documentclass[useAMS,usenatbib]{mn2e}
\usepackage{graphicx}
\usepackage{natbib}
\usepackage{subfigure}
\usepackage{placeins}

\input{psfig.sty}

\newcommand{\AIPS}{{$\cal AIPS\/$}}

\def\herschel{{\it Herschel}}

\def\l1.4{$L_{\rm 1.4GHz}$}
\def\s1.4{$S_{\rm 1.4GHz}$}

\def\kms{km\,s$^{-1}$ }

\def\jonezero{$J\!=\!1\!-\!0$}
\def\jthreetwo{$J\!=\!3\!-\!2$}

\def\gs{\mathrel{\raise0.35ex\hbox{$\scriptstyle >$}\kern-0.6em
\lower0.40ex\hbox{{$\scriptstyle \sim$}}}}
\def\ls{\mathrel{\raise0.35ex\hbox{$\scriptstyle <$}\kern-0.6em
\lower0.40ex\hbox{{$\scriptstyle \sim$}}}}
\def\m@th{\mathsurround=0pt }
\def\eqalign#1{\null\,\vcenter{\openup1\jot \m@th
 \ialign{\strut\hfil$\displaystyle{##}$&$\displaystyle{{}##}$\hfil
 \crcr#1\crcr}}\,}
\title[VLA imaging of ionised and molecular gas in SMGs]{VLA imaging of \co\ \jonezero\ and free-free emission in lensed submillimetre galaxies}

\author[A.\,P.\ Thomson \etal]{A.\,P.\ Thomson,$^{\! 1}$\thanks{E-mail: at@roe.ac.uk}
  R.\,J.\ Ivison,$^{\! 1,2}$
  Ian Smail,$^{\! 3}$
  A.\,M.\ Swinbank,$^{\! 3}$
  A.\ Weiss,$^{\! 4}$
  J.-P.\ Kneib,$^{\! 5}$ \and
  P.\,P.\ Papadopoulos,$^{\! 4}$
  A.\,J.\ Baker,$^{\! 6}$
  C.\,E.\ Sharon$^{6}$ and
  G.\,A.\ van Moorsel$^{7}$
  \vspace*{1mm}\\
  $^{1}$Institute for Astronomy, University of Edinburgh, Blackford Hill, Edinburgh EH9 3HJ\\
  $^{2}$UK Astronomy Technology Centre, Science and Technology Facilities Council,
          Royal Observatory, Blackford Hill, Edinburgh EH9 3HJ\\
  $^{3}$Institute for Computational Cosmology, Durham University, South Road, Durham DH1 3LE\\
  $^{4}$Max-Planck-Institut f\"{u}r Radioastronomie, Auf dem H\"{u}gel 69, 53121, Bonn, Germany\\
  $^{5}$Laboratoire d'Astrophysique de Marseille, CNRS-Universit\'{e}
          Aix-Marseille, 13388 Marseille Cedex 13, France\\
  $^{6}$Department of Physics and Astronomy, Rutgers, The State of New
           Jersey, 136 Frelinghuysen Road, Piscataway, NJ 08854-8019, USA\\
  $^{7}$National Radio Astronomy Observatory, 1003 Lopezville Road, Socorro, NM 87801, USA}

\date{Accepted 2012 June 24. Received 2012 June 22; in original form in 2012 May 22}
\pagerange{\pageref{firstpage}--\pageref{lastpage}} \pubyear{2012}

\begin{document}
\newcommand{\co}{$^{12}$CO}
\newcommand{\etal}{et~al.}
\newcommand{\msol}{\,\textrm{M}_{\odot}}                
\newcommand{\lsol}{\,\textrm{L}_{\odot}}                
\newcommand{\hst}{\textit{HST}}
\newcommand{\mdyn}{$M_{\rm dyn}$}
\newcommand{\mstar}{M$_{*}$}
\newcommand{\sfr}{M$_{\odot}$\,yr$^{-1}$}

\maketitle
\label{firstpage}
\begin{abstract}
  We present a study using the Karl G.\ Jansky Very Large Array (VLA)
  of \co\ \jonezero\ emission in three strongly lensed
  submillimetre-selected galaxies (SMM\,J16359, SMM\,J14009 and
  SMM\,J02399) at $z=2.5$--2.9. These galaxies span $L_{\rm IR} =
  10^{11-13}$\,L$_{\odot}$, offering an opportunity to compare the
  interstellar medium of LIRGs and ULIRGs at high redshift.  We
  estimate molecular gas masses in the range $2-40 \times
  10^{9}$\,M$_{\odot}$ using a method that assumes canonical
  underlying brightness temperature ($T_{\rm b}$) ratios for
  star-forming and non-star-forming gas phases and a maximal
  star-formation efficiency. A more simplistic method -- using $X_{\rm
    CO} = 0.8$ and the measured $T_{\rm b}$ ratios -- yields gas
  masses twice as high. In SMM\,J14009 we find $L'_{\rm CO3-2}/L'_{\rm
    CO1-0} = 0.95 \pm 0.12$, indicative of warm, star-forming gas,
  possibly influenced by the central active galactic nucleus (AGN). We
  set a gas mass limit of $3\sigma<6\times 10^8$\,M$_{\odot}$ for the
  Lyman-break galaxy, A2218 \#384, located in the same field as
  SMM\,J16359 at $z=2.515$. Finally, we use the rest-frame
  $\sim$115-GHz free-free flux densities for SMM\,J14009 and
  SMM\,J02399 -- measurements tied directly to the photoionisation
  rate of massive stars, and made possible by VLA's bandwidth -- to
  estimate star-formation rates (SFRs) of
  400--600\,$\msol$\,yr$^{-1}$, and to estimate the fraction of
  $L_{\rm IR}$ due to AGN.
\end{abstract}


\section{Introduction}\label{sec:intro}

Observations of the interstellar medium (ISM) in high-redshift
galaxies can provide insight into the process of star formation within
these systems.  Although the bulk of the molecular gas -- the fuel for
future star formation -- is in the form of H$_2$, its lack of a net
dipole moment prevents H$_2$ from radiating strongly under normal ISM
conditions. However, this gas can be detected indirectly due to the
collisional excitation it induces in heteronuclear and near-ubiquitous
\co\ -- the second most abundant molecule in the ISM -- which emits
electromagnetic radiation when its angular momentum quantum number
changes by $\hbar$. The low dipole moment of \co\ enables its
excitation in regions of low density, $n_{\rm crit} \geq
10^2$\,cm$^{-3}$, making it an ideal tracer of the bulk of the
molecular gas reservoir, allowing us to determine its properties in a
manner unbiased by the patchy dust extinction which can lead to
misleading views of velocity fields and galaxy morphologies via
optical/near-infrared studies.

Since the earliest detections of \co\ in high-redshift galaxies
\citep[e.g.][]{brown91, solomon92}, much effort has been put into
discerning the mass, extent and physical properties of the H$_2$ gas
in galaxies across cosmic time; this offers a snapshot of their
evolutionary state and narrows down the possibilities for their likely
descendents.  Until now, available technology has limited the study of
\co\ at high redshift to relatively bright, mid-$J$ lines (\co\
\jthreetwo\ and above), which are shifted into broad, clean
atmospheric windows such as that at 3\,mm. The excitation requirements
of these lines render such observations insensitive to any cool,
extended gas component, wherein much of the gas is routinely found
\citep[e.g.][]{papadopoulos99, weiss05, bothwell12}. For local
galaxies the intrinsically fainter \jonezero\ transition occupies the
3-mm atmospheric window and mid-$J$ lines are more difficult to study,
requiring high-frequency receivers and excellent atmospheric
conditions. Efforts to conduct comparative studies at low and high
redshift have therefore been few and far between, and the relationship
between high-redshift galaxies and their myriad possible present-day
descendants is not well understood.

Gravitational amplification -- either by chance, line-of-sight
foreground galaxies \citep[e.g.][]{rowan-robinson91, graham95,
  negrello10} or by deliberate exploitation of massive, foreground
clusters \citep[e.g.][]{kneib10} -- has provided samples that have
formed the basis for much of our understanding of the distant
starbursts that dominate the cosmic far-infrared/submm
background. Usually buried in the confusion noise, these faint
galaxies become relatively easy to detect and spatially resolve by
virtue of the magnification they experience. Such an approach led to
the discovery of submillimetre-selected galaxies \citep[SMGs --
][]{smail97}, to the relatively straightforward identification of
their radio and optical counterparts, acquisition of their redshifts
\citep[e.g.][]{ivison98} and their detection in mid-$J$ \co\
\citep[e.g.][]{frayer98}. The redshift distribution of SMGs is broad,
centred at $z \sim 2.3$ \citep{chapman05}, and their inferred stellar
masses ($\sim 10^{10-11}$\,$\msol$) and star-formation rates (SFRs;
$\sim 10^{3}$\,\sfr) make them among the most massive and active
galaxies at this epoch \citep[e.g.][]{hainline11}.

Here, we follow this familiar pattern. We exploit the recent upgrade
to the National Radio Astronomy Observatory's (NRAO's\footnotemark[1])
Karl G.\ Jansky Very Large Array (VLA), which includes the provision
of Ka-band receivers (26.5--40\,GHz), to study the redshifted \co\
\jonezero\ emission from three lensed SMGs: (i) SMM\,J14009+0252 lying
at $z=2.9344$ \citep{ivison00, weiss09}, lensed by Abell\,1835 and one
of the brightest SMGs detected in the SCUBA Cluster Lens Survey
\citep[SCLS; ][]{smail02} (ii) SMM\,J16359+6612, a bright SMG a
$z=2.5168$ \citep{kneib04, kneib05} and lensed into three images by
the Abell\,2218 cluster, and (iii) SMM\,J02399--0136 at $z=2.808$
\citep{ivison98}, the first SCUBA galaxy and the brightest of the SCLS
galaxies, lensed by Abell\,370. We also search for \co\ \jonezero\
emission in (iv) the Lyman-break galaxy (LBG) A\,2218 \#284, located
22.1\,arcsec from the centre of the lensing cluster in the same field
as SMM\,J16359 \citep{ebbels96}.

The optical spectrum of SMM\,J02399, coupled with its extreme
luminosity and emission-line widths ({\sc fwzi} $\sim
1,000$\,km\,s$^{-1}$) have long been cited as evidence that this
source contains a type-2 active galactic nucleus \citep[AGN;
][]{ivison98}. SMM\,J14009 too has long been suspected to host a
dust-enshrouded AGN on the basis of the strength of its radio emission
($S_{\rm 1.4\,GHz} = 529 \pm 30\mu$Jy) and compactness \citep[Biggs
\etal\ \textit{in prep.} ]{ivison00}.

Due to the effects of lensing, we benefit from a boost in flux density
and apparent size of our targets, surface brightness being
conserved. Excellent lens models exist for Abell 370, 1835 and 2218
and multi-wavelength imaging has constrained the ages, stellar masses
and SFRs of our targets, as well as the mass of warm, star-forming gas
as traced by, e.g., \co\ \jthreetwo. Observations of another SMG
lensed by A\,1835, SMM\,J14011+0252 at $z=2.5652$, will be discussed
by Sharon \etal\ \textit{in prep}.  \footnotetext[1]{NRAO is operated
  by Associated Universities Inc., under a cooperative agreement with
  the National Science Foundation.}

Throughout the paper we use a cosmology with $H_0 =
71$\,km\,s$^{-1}$\,Mpc$^{-1}$, $\Omega_{\rm m}=0.27$, $\Omega_\Lambda
= 0.73$. which gives an average angular scale of
8.1\,kpc\,arcsec$^{-1}$ for our sample.

%
%
\begin{table*}\label{tab:sample}
  \centering
  \caption{VLA sample and observing log.}
  \begin{tabular}{lccccccl}
    \hline
    \multicolumn{1}{l}{Target} &
    \multicolumn{1}{c}{Magnifi-} &
    \multicolumn{1}{c}{$z$} &
    \multicolumn{1}{c}{$\Delta\nu_{\rm obs}$} &
    \multicolumn{1}{c}{Configu-} &
    \multicolumn{1}{c}{Observing dates} &
    \multicolumn{1}{l}{Reference} \\
    \multicolumn{1}{l}{} &
    \multicolumn{1}{c}{-ication} &
    \multicolumn{1}{c}{} &
    \multicolumn{1}{c}{(GHz)} &
    \multicolumn{1}{c}{-ration} &
    \multicolumn{1}{c}{} &
    \multicolumn{1}{c}{} \\
     \hline
SMM\,J14009&1.5 &2.9344&29.23  &DnC&Sep 25, 29, 30 (2010),           &\citet{weiss09} \\
          &      &     &--29.36&CnB&Jan 21, 24, 26, 27 (2011)\\
SMM\,J16359&14, 22, 9$^a$  &2.5168&32.69  &D  &Apr 12, 14, 20, May 05, 20 (2010),&\citet{kneib04}   \\
           &               &      &--32.93&C  &Dec 31, Jan 04, 06, 10, 11 (2011)& \\
SMM\,J02399&2.45             &2.808&29.73     &D  &Sep 27, Oct 26 $\times2$,    & \citet{genzel03}    \\
		&      &&30.69&  &Nov 01 $\times 2$ (2011) & \\
	\hline
\end{tabular}
{\small
\\Note: $^a$Images A, B, C, respectively.}
\end{table*}

\section{Observations and data reduction}\label{sec:data-reduction}

\subsection{VLA observations}

Observations of SMM\,J14009 and SMM\,J16359 were carried out in blocks
of 2--4\,hr during excellent weather conditions between 2010 April and
2011 January (Project IDs: AI139 and AI142). Over this Open Shared
Risk Observing period the available bandwidth from the new Wideband
Interferometric Digital ARchitecture (WIDAR) correlator consisted of
two sub-band pairs, each with $64\times 2$-MHz dual-polarisation
channels, and the number of useful Ka receivers roughly doubled, from
$\sim10$ to $\sim20$. Due to the different redshifts of the targets,
different tuning strategies had to be employed in each case: for
SMM\,J16359 we used both sub-band pairs with the frequency of the
redshifted \co\ \jonezero\ line \citep[$\nu_{\rm rest} =
115.27120256$\,GHz -- ][]{morton94} placed in channel 40 of the lower
sub-band and a 10-channel overlap between the two sub-bands to
mitigate the effects of noise in the edge channels. This configuration
gave velocity coverage and resolution of $\sim$2,000\,\kms\ and
$\sim$18\,\kms, respectively. For the higher-redshift SMM\,J14009, the
redshifted emission was accessible to only one of the two sub-bands
and the line was placed in its centre providing $\sim$1,300\,\kms\
coverage, the other sub-band being tuned to $\nu_{\rm obs}$ =
32.4\,GHz in order to constrain the $\sim$130-GHz radio continuum.

By the time SMM\,J02399 was observed during 2011 September through
November (Project ID: AT400), the OSRO capabilities had been upgraded
to include two independently tunable output pairs of eight sub-bands
each, with $128\times 1$-MHz dual-polarisation channels per sub-band,
giving a total available bandwidth of 2,048\,MHz. Again, however, the
redshift of SMM\,J02399 could be reached by only the BD output pair,
giving $\sim$20,000\,\kms\ of coverage and $\sim$10\,\kms\
resolution. The 8 sub-bands of output pair AC were tuned continuously
at the lowest possible central frequency, 32.52\,GHz, to constrain the
radio continuum emission from this source.

For SMM\,J16359 we obtained approximately 19 and 14\,hr of useful data
in the VLA's D and C configurations; for the equatorial target
SMMJ\,14009 we employed the DnC and CnB configurations (with their
longer northern arms), acquiring 8 and 16\,hr, respectively. 10\,hr of
on-source D configuration observations were performed for SMM\,J02399,
though an unfortunate technical problem with the correlator resulted
in only 1/3 of the observed data being recorded.

Antenna pointing was checked using the C-band receivers every
60--90\,min, and immediately prior to scans of the flux calibrator
(3C\,286 for SMM\,J14009 and SMM\,J16359; 3C\,48 for SMM\,J02399). We
tracked amplitude and phase with brief scans of J1642+6856 (for
SMM\,J16359), J1354$-$0206 (for SMM\,J14009) and J0239$-$0234 (for
SMM\,J02399), every 5\,min. These data were also used to determine the
bandpass.

Data were flagged, calibrated and imaged using standard \AIPS\
procedures, as outlined by \citet{ivison11}.

We constructed data cubes for each source using the {\sc clean}
algorithm with 6-MHz ($\sim50$-\kms) resolution by binning the native
VLA channels to boost signal-to-noise. We used natural weighting and
Gaussian $uv$ tapers to check for any flux which may have been
resolved out in the high-resolution images but found no evidence for
significant missing flux in any of our sources. The resulting
synthesised beam sizes are $0.88\times 0.72$\,arcsec$^{2}$ with
position angles (PAs) $125^{\circ}$ for SMM\,J14009,
$1.07\times0.72$\,arcsec$^{2}$ at PA $22^{\circ}$ for SMM\,J16359 and
$3.08 \times 2.56$\,arcsec$^{2}$ at PA $160^{\circ}$ for SMM\,J02399;
the integrated maps (Fig.~1) reach down to a noise levels of 20, 11
and 20\,$\mu$Jy\,beam$^{-1}$, respectively.

Spectra (see Fig.~2) were extracted from the data cubes with the
\AIPS\ task, {\sc ispec}, which sums the flux density within a box
around the source for each plane in the cube. Error spectra were
produced by extracting five off-source spectra and calculating the
standard deviations among these five sky spectra for each spectral
channel. The mean noise values per spectral channel for SMM\,J14009,
SMM\,J16359 and SMM\,J02399 were 0.09, 0.17 and
0.13\,mJy\,beam$^{-1}$, respectively.

Line fluxes are measured by integrating the emission spectra across
the width of each \co\ line and then checked via the equivalent
integrated image. In determining both the observational and derived
source properties, we treat the three images of SMM\,J16359 as
individual sources and then co-add the results to derive
magnification-weighted averages where physically appropriate.

\subsection{Infrared luminosities}\label{sec:lir}

Throughout the paper, we make reference to infrared properties of our
sources, including calculations of the gas masses and SFR derived from
rest-frame 8--1,000\,$\mu$m luminosities, $L_{\rm IR}$. Here, we
present the measurements used to derive these quantities.

We use archival data from \herschel\ \citep{pilbratt10} to measure
$S_{250 \mu m}$, $S_{350\mu m}$ and $S_{500 \mu m}$, together with
published SCUBA flux densities at 750, 850 and 1350\,$\mu$m. We then
exploit the SED library of \citet[ hereafter CE01]{chary01} in order
to determine $L_{\rm IR}$ for each galaxy.

For SMM\,J02399 we use the mid-IR spectrum of \citet{Lutz05}, with
data covering $\lambda_{\rm obs} \geq 750 \mu$m taken from
\citet{ivison98}. In Fig.~\ref{fig:seds} we see the excellent
agreement between the data and the best-fit SED template. The derived
value of $L_{\rm IR} = (80.7 \pm 8.1) \times 10^{11}$\,L$_{\odot}$ is
30 per cent lower than the value reported in \citet{ivison98}.

For SMM\,J16359 we use published flux densities for the three images
at 24 and 70\,$\mu$m from \citet{kneib04}. The three images are
resolved by \herschel\ up to 350\,$\mu$m but become blended at longer
wavelengths. Where the three images are blended, we divide up the
observed flux in proportion to the known magnification factors. We
derive an estimate of $L_{\rm IR}$ for each of the three images of this
source and combine our three estimates to derive the
magnification-weighted mean $L_{\rm IR} = (6.4 \pm 1.0) \times
10^{11}$\,L$_{\odot}$. This value is consistent with the results of
\citet{finkelstein11}, who determined $L_{\rm IR} = 7.0 \times
10^{11}$\,L$_{\odot}$ (for image B) using a similar SED-fitting
method.

Similarly, for SMM\,J14009 we rely on the 24-$\mu$m measurement from
\citet{hempel08} to supplement the archival \textit{Herschel} and
SCUBA data, and compute a lensing-corrected $L_{\rm IR} = (160 \pm 24)
\times 10^{11}$\,L$_{\odot}$.

The best-fit CE01 SEDs for each galaxy are shown with the
corresponding photometry in Fig.~3.

\subsection{Radio continuum data}\label{subsect:radio-continuum}

The VLA tuning strategies employed to observe \co\ \jonezero\ in for
SMM\,J14009 and SMM\,J02399 enabled us to obtain detections of the
$\sim115$-GHz continuum emission from these galaxies. These data were
reduced in parallel with and according to the same recipe as the \co\
spectral-line data. We do not have continuum data at this frequency
for SMM\,J16359, owing to the different tuning strategy employed at
the VLA for this source. We detect continuum flux densities of
$S_{\rm 32\,GHz} = 57 \pm 25$\,$\mu$Jy for SMM\,J02399 and $S_{\rm
  32\,GHz} = 42 \pm 20$\,$\mu$Jy for SMM\,J14009. 1.4-GHz flux
densities of $S_{\rm 1.4\,GHz} = 526\pm50$ and $529\pm30$\,$\mu$Jy
were determined in \citet{ivison98} and \citet{ivison00} for
SMM\,J02399 and SMM\,J14009, respectively. The VLA science archive was
queried for continuum data at 5- and 8-GHz as well, with flux
densities reported in Table.~4. The radio continuum data are plotted
with the IR SEDs in Fig.~3.

\section{Analysis and results}\label{sec:analysis-results}

\subsection{\co\ morphologies and spectra}\label{subsect:morphologies}

\begin{figure*}\label{fig:co-images}
\centerline{\psfig{file=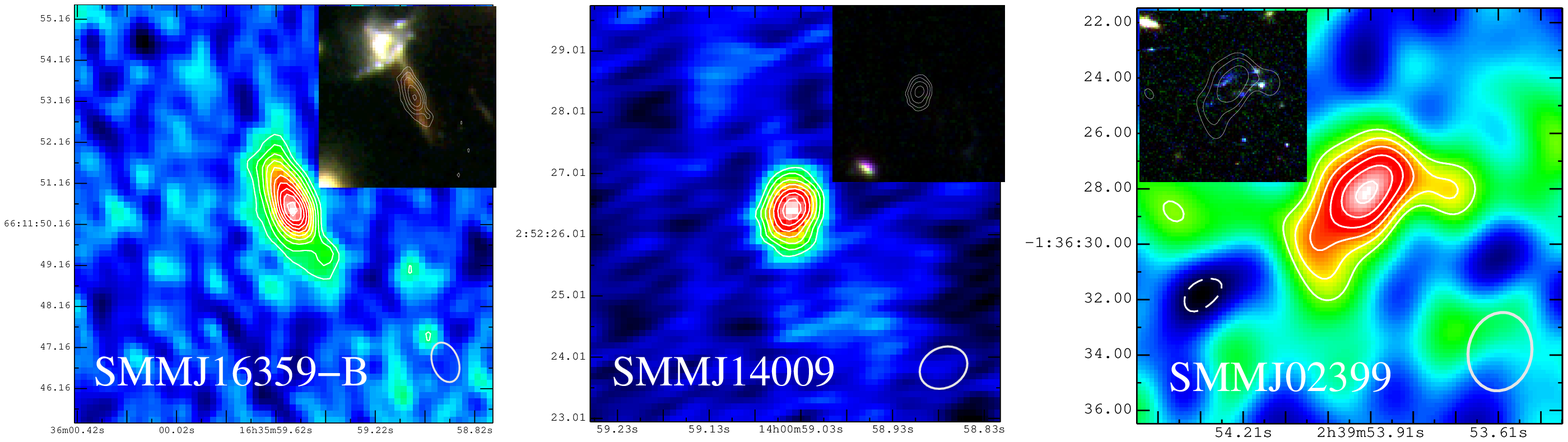,width=17cm}}
\caption{Integrated \co\ \jonezero\ maps of our three targets,
  SMM\,J16359 (10 arcsec/side), SMM\,J14009 (10 arcsec/side) and
  SMM\,J02399 (15 arcsec/side). For SMM\,J16359 -- a triply imaged
  source -- we show only the brightest image, SMM\,J16359-B as defined
  by \citet{kneib04}. \textit{Main images:} \co\ \jonezero\ contours
  spaced at $-$3, 3, 4, 5...$\times\sigma$, set on top of colour
  representations of the same data. The VLA beam used to produce each
  map is shown in the bottom-right corner. \textit{Insets:} \co\
  \jonezero\ contours spaced at $-$3, 3,
  $\sqrt{2}\times$3...$\times\sigma$ and in steps of
  $\sqrt{2}\times\sigma$ thereafter, set on top of three-colour \hst\
  images of the same field of view, in order to explore the
  relationship between the SMG and any optical counterpart there may
  be. For SMM\,J14009 we detect no optical counterpart to the \co\
  emission.}
\end{figure*}

\begin{figure*}\label{fig:co-spectra}
\centerline{\psfig{file=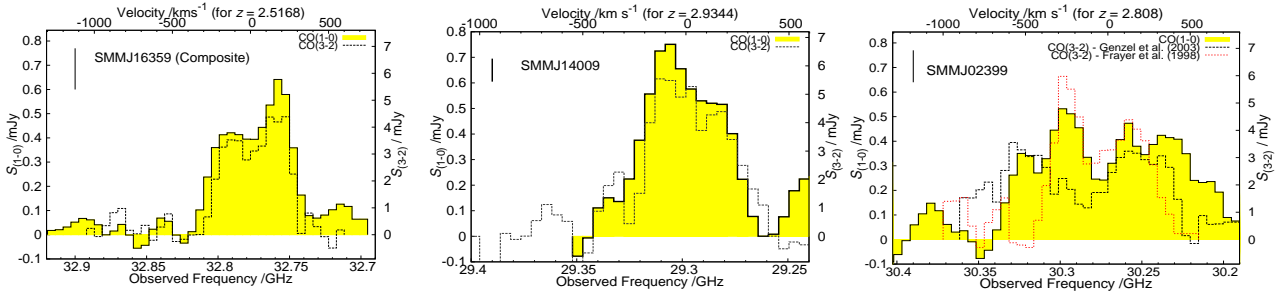,width=17cm}}
\caption{\co\ \jonezero\ spectra (yellow histograms) and reference
  \jthreetwo\ spectra (dotted lines) for our three targets. The
  left-hand scale represents \jonezero\ flux density, and the
  right-hand side shows the corresponding $J=3$--2 flux density,
  scaled by $9^{-1}$ so as to be on the same Rayleigh-Jeans $T_{\rm
    b}$ scale. The zero-LSR redshifts used to determine the relative
  velocities for each source are quoted in the figure, and typical
  error bars are shown in the top-left-hand corner of each plot. The
  spectrum shown for SMM\,J16359 is the magnification-weighted mean of
  the spectra of the three images of this source, SMM\,J16359-A, -B
  and -C as defined in \citet{kneib04}.}
\end{figure*}

\begin{figure*}
\centerline{\psfig{file=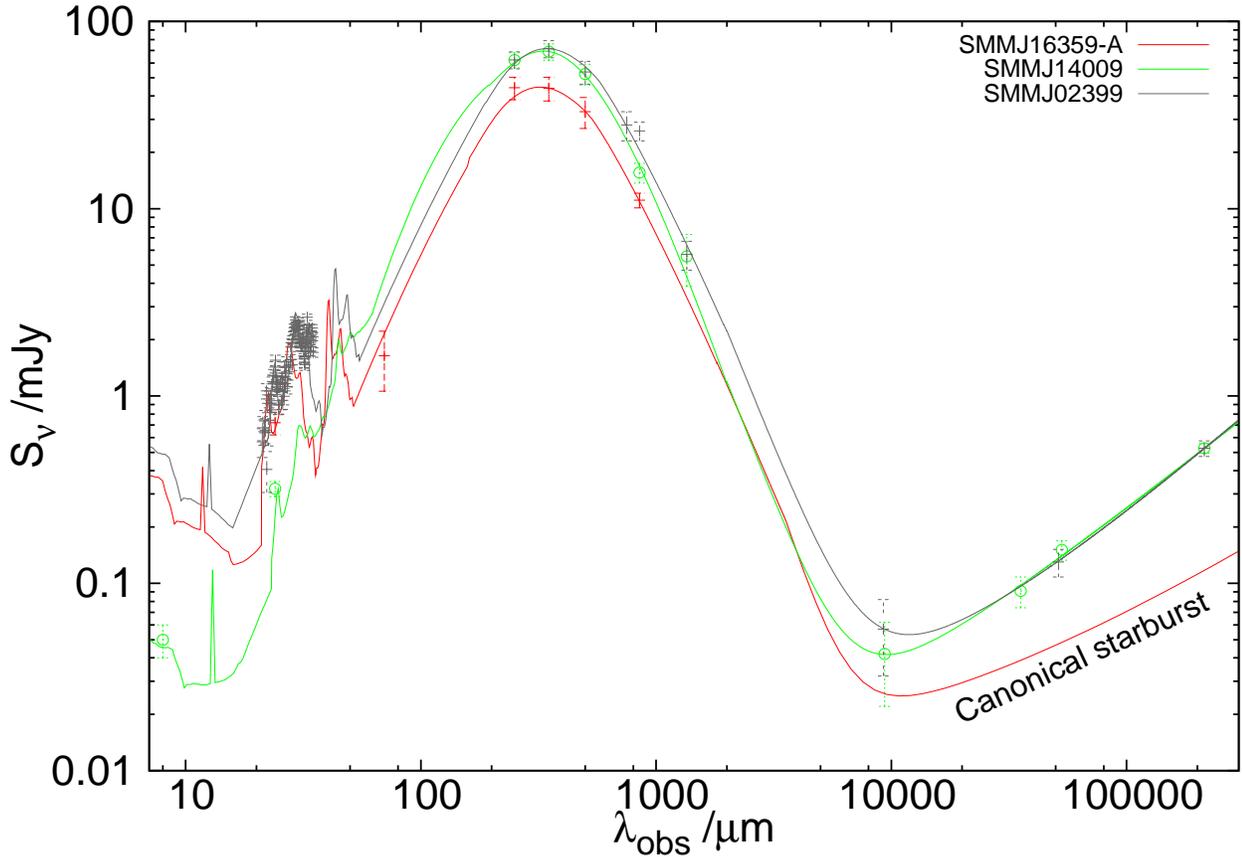,width=17cm}}
\caption{Observed-frame infrared SEDs for all of our sources,
  uncorrected for lensing, including data (\textit{symbols: see
    legend}) and the corresponding best-fit CE01 SED. We include only
  data up to $\lambda_{\rm obs} = 1350\,\mu$m in the fit. For clarity,
  we plot the SED of only one image of SMM\,J16359, and plot the
  best-fit CE01 template out to radio wavelengths, demonstrating the
  ``canonical starburst'' radio SED. For the other two galaxies, we
  truncate the CE01 SED at 2,000\,$\mu$m and plot our modelled radio
  SED, constrained by our measurement of the Ka-band continuum flux
  and previously published results at longer wavelengths
  (\S\ref{subsect:radio-continuum}, \S\ref{sect:radio-continuum}) and
  interpret the steepness of the resulting spectra as being consistent
  with the known AGN hosted within these galaxies.}
\label{fig:seds}
\end{figure*}

We summarise the source properties in Table~2. All sources are
detected in \co\ \jonezero\, with signal-to-noise of 13 for
SMM\,J16359-B (the brightest of three images of this multiply-lensed
source), 17 for SMM\,J14009 and 7.5 for SMM\,J02399. The flux-weighted
\co\ \jonezero\ redshifts -- summarised in Table~2 -- are derived by
fitting a single Gaussian component to each \co\ \jonezero\, spectrum
and are consistent with the reference \co\ \jthreetwo\ redshifts to
four significant figures. The spectrum shown for SMM\,J16359 is a
magnification-weighted average of the spectra from each image of this
source. The uncertainties in the spectral fits translate to velocity
uncertainties that are typically of order $\Delta V=10$--15\,\kms,
except in the case of SMM\,J02399 where the comparatively low
signal-to-noise hampers our ability to constrain $z_{\rm CO}$ and
results in $\Delta V\sim50$\,\kms. For each source we see broad lines
with full width at zero intensity ({\sc{fwzi}}) of
$\sim$500--1,000\,\kms.

We compare the line profiles and maps between \jonezero\ and
\jthreetwo\ observations. The astrometry is consistent. For
SMM\,J16359 and SMM\,J14009 we see good agreement between the low-$J$
and mid-$J$ \co\ line profiles. In the former, we detect double-peaked
\jonezero\ line profiles for each of the three images, separated by
$\sim$300\,\kms, as did \citet{kneib05} in their study of the
\jthreetwo\ emission.  SMM\,J02399 displays a significant offset in
velocity between the measured \jonezero\ emission and the \jthreetwo\
spectrum of \citet{genzel03}, $\sim$100\,\kms, roughly twice that
which could be ascribed to uncertainties in the Gaussian
fits. However, the \jonezero\ and \jthreetwo-derived redshifts are
consistent to four significant figures ($z = 2.808$). It is difficult
to envisage any plausible physical origin for these offsets: in an
effort to track down any instrumental contribution to the frequency
offsets, continuous-wave tones were injected into the Ka-band
receivers at frequencies close to those used for the science
observations, but no offsets were observed between the input and
observed tones (Emmanuel Momjian, private communication).

We find a weighted mean velocity width ratio of $\sigma_{(1-0)/(3-2)}
= 1.3\pm 0.1$ for the three images of SMM\,J16359, consistent with the
findings of \citet{ivison11} for a sample of four unlensed SMGs. For
the two more AGN-like galaxies, SMM\,J14009 and SMM\,J02399, we detect
velocity width ratios consistent with unity ($\sigma_{(1-0)/(3-2)} =
1.06 \pm 0.08$ and $1.05 \pm 0.26$ respectively). 
%
%
\begin{table*}
  \centering
  \caption{Observed properties.}
  \begin{tabular}{lcccccccc}
    \hline
    \multicolumn{1}{l}{Target} &
    \multicolumn{1}{c}{$z_{\rm CO(1-0)}$} &
    \multicolumn{1}{c}{$I_{\rm CO(1-0)}$} &
    \multicolumn{1}{c}{$I_{\rm CO(3-2)}$} &
    \multicolumn{1}{c}{$r_{\rm 3-2/1-0}$} &
    \multicolumn{1}{c}{$\sigma_{\rm CO(1-0)}$} &
    \multicolumn{1}{c}{$\sigma_{\rm CO(1-0)/CO(3-2)}$} &
    \multicolumn{1}{c}{{\sc{fwzi}}} &
    \multicolumn{1}{c}{R} \\
    \multicolumn{1}{l}{} &
    \multicolumn{1}{c}{} &
    \multicolumn{1}{c}{(Jy\,km\,s$^{-1}$)} &
    \multicolumn{1}{c}{(Jy\,km\,s$^{-1}$)} &
    \multicolumn{1}{c}{} &
    \multicolumn{1}{c}{(km\,s$^{-1}$)} &
    \multicolumn{1}{c}{(km\,s$^{-1}$)} &
    \multicolumn{1}{c}{(km\,s$^{-1}$)} &
    \multicolumn{1}{c}{kpc} \\
     \hline
SMM\,J14009&$2.9332 \pm 0.0001$  &$0.31 \pm 0.02$&$2.70 \pm 0.30 $  &$0.95 \pm 0.12$&$175 \pm 10$&$1.06 \pm 0.08$&850&$1.2 \pm 0.4$\\
SMM\,J02399&$2.8083 \pm 0.0007$    &$0.60 \pm 0.12$&$3.10 \pm 0.40 $  &$0.58 \pm 0.14$&$350 \pm 90$&$1.05 \pm 0.26$           &1300&$26 \pm 4$\\
SMM\,J16359-A&$2.5174 \pm 0.0002$  &$0.22 \pm 0.04$&$1.67 \pm 0.13 $  &$0.83 \pm 0.15$&$195 \pm 15$&$1.10 \pm 0.09$&710&$<1$\\
SMM\,J16359-B&$2.5173 \pm 0.0001$  &$0.40 \pm 0.04$&$2.50 \pm 0.12 $  &$0.70 \pm 0.22$&$190 \pm 10$&$1.26 \pm 0.07$&780&$0.6 \pm 0.3$\\
SMM\,J16359-C&$2.5179 \pm 0.0002$  &$0.30 \pm 0.09$&$1.58 \pm 0.17 $  &$0.59 \pm 0.20$&$305 \pm 15$&$1.81 \pm 0.09$&670&$4 \pm 2$\\
    \hline
  \end{tabular}
\end{table*}

\subsection{Line luminosities and derived total gas masses}\label{subsec:gas-masses}

%
%
\begin{table*}\label{tab:derived-properties}
  \centering
  \caption{Lensing-corrected properties}
  \begin{tabular}{lccccccc}
    \hline
    \multicolumn{1}{l}{Target} &
    \multicolumn{1}{c}{$L^{'}_{\rm CO(1-0)}$} &
    \multicolumn{1}{c}{$M_{\rm gas, \alpha}$} &
    \multicolumn{1}{c}{$m_{\rm CQ/SF}$} &
    \multicolumn{1}{c}{$L_{\rm IR}^a$} &
    \multicolumn{1}{c}{$M_{\rm gas, I11}$} &
    \multicolumn{1}{c}{$M_{\rm dyn}$} \\
    \multicolumn{1}{l}{} &
    \multicolumn{1}{c}{($\times 10^9$ K\,km\,s$^{-1}$\,pc$^{2}$)} &
    \multicolumn{1}{c}{($\times 10^9$ M$_{\odot}$)} &
    \multicolumn{1}{c}{} &
    \multicolumn{1}{c}{($\times 10^{11}$ L$_{\odot}$)} &
    \multicolumn{1}{c}{($\times 10^9$ M$_{\odot}$)} &
    \multicolumn{1}{c}{($\times 10^9$ M$_{\odot}$)}\\
    \hline
SMM\,J14009&$83 \pm 6$  &$67 \pm 4$&$0.09 \pm 0.02$&$160 \pm 16$&$35 \pm 7$&$18 \pm 4$\\
SMM\,J02399  &$96.6 \pm 1.9$  &$77 \pm 15$&$1.5 \pm 0.5$&$81 \pm 8$&$41 \pm 14$&$1560 \pm 550$\\
SMM\,J16359-A&$4.9 \pm 0.8$  &$3.9 \pm 0.6$&$0.33 \pm 0.09$&$7.1 \pm 0.7$&$1.3 \pm 0.4$&$<19$\\
SMM\,J16359-B&$5.5 \pm 0.6$  &$4.4 \pm 0.5$&$0.8 \pm 0.3$&$6.2 \pm 0.6$&$2.8 \pm 1.3$&$11 \pm 4$\\
SMM\,J16359-C&$10 \pm 3$   &$8 \pm 2$&$1.4 \pm 0.7$&$5.7 \pm 0.6$&$1.5 \pm 0.7$&$180 \pm 70$\\
A2218 \#384&$3\sigma<0.8$ & $3\sigma<0.6$ & -- & -- & -- & --\\
    \hline
  \end{tabular}
  {\small
    \\ Note: $^a$Calculations for $L_{\rm IR}$ discussed in \S\ref{sec:lir}. All luminosities have been corrected for the known magnification factors.}
\end{table*}

Traditionally, attempts to determine the molecular gas mass of distant
galaxies via \co\ emission have been subject to two main sources of
uncertainty: (1) where mid-$J$ lines are observed, a brightness
temperature ratio, $r_{(3-2)/(1-0)} = 1$ has typically been assumed to
convert the \textit{measured} \co\ \jthreetwo\ flux to an
\textit{inferred} \jonezero\ flux, and (2) the ``X-factor'', $X_{\rm
  CO} = M({\rm H}_2)/L'_{\rm CO(1-0)}$ \citep{downes98} which is used
to convert the observed or inferred \co\ \jonezero\ luminosity to a
total molecular gas mass. As a number of recent papers have shown
\citep[e.g.][etc.]{harris10, ivison11, danielson11}, observed
$r_{(3-2)/(1-0)}$ ratios in SMGs are typically well below unity. We
can thus improve our gas mass estimates by observing the \co\ line
which traces the bulk of the molecular gas reservoir, but we must
still worry about the order of magnitude range spanned by $X_{\rm CO}$
depending on the assumed state of the underlying ISM: $X_{\rm CO} \sim
5$ in optically-thick giant molecular clouds (GMCs) where the gas is
reducible to ensembles of self-gravitating units, whereas $X_{\rm CO}
\sim 0.8$ in the more extreme environments of local ultra luminous
infrared galaxies (ULIRGs), where the ISM forms a smooth, continuous
medium. This latter value is typically adopted for SMGs.

The intrinsic \co\ \jonezero\ line luminosity of an unlensed galaxy is
the observed line flux integrated across the surface area and velocity
range of the line: $L'_{\rm CO} = \int_{\Delta V} \int_{A_{s}}\,
S_{\nu} \,dA \,dV$\,{\sc k}\,km\,s$^{-1}$\,pc$^{2}$, or:

\begin{center}
\begin{equation}\label{eq:L_CO}
L'_{\rm CO(1-0)} = 3.25 \times 10^{7} \biggl[\frac{D_{\rm L}^{2}({\rm Mpc})}{1+z} \biggr] \biggl(\frac{\nu_{\rm rest}}{\rm GHz} \biggr) ^{-2}
\end{equation}
$\times \biggl[ \frac{\int_{\Delta V} S_{\nu}d\nu}{\rm Jy\,km\,s^{-1}\,pc^{2}} \biggr] {\rm K\,km\,s^{-1}\,pc^{2}}$
\end{center}

Having made an appropriate correction for magnification effects, we
then invoke the commonly-used CO--H$_{2}$ gas mass conversion factor
of \citet{downes98} of $X_{\rm CO} \sim 0.8$\,M$_{\odot}$\,({\sc
  k}\,km\,s$^{-1}$\,pc$^{2}$)$^{-1}$ to derive total gas masses
(Table~3).

Combining our \co\ \jonezero\ data with existing \jthreetwo\ data from
\citet[ SMM\,J16359]{kneib05}, \citet[ SMM\,J14009]{weiss09} and
\citet[ SMM\,J02399]{genzel03} enables us to derive a mean
magnification-weighted $T_{\rm b}$ ratio of $r_{(3-2)/(1-0)} = 0.72
\pm 0.12$ for the three images of SMM\,J16359, $r_{(3-2)/(1-0)} = 0.95
\pm 0.25$ for SMM\,J14009 and $r_{(3-2)/(1-0)} = 0.58 \pm 0.14$ for
SMM\,J02399: SMGs on average experience \textit{sub-thermal}
excitation. We can then compare the gas mass derived above ($M_{\rm
  gas, \alpha}$) with a gas mass derived according to the prescription
of \citet{ivison11}, denoted $M_{\rm gas, I11}$, in which a lower
limit for the mass of actively star-forming gas ($M_{\rm SF}$) is
computed on the assumption that SMGs form stars at an appreciable
fraction of Eddington-Limited star-formation efficiency
\citep[SFE$_{\rm max} = L_{\rm IR}/M_{\rm SF} \leq
500$\,L$_{\odot}$\,M$_{\odot}^{-1}$ where $L_{\rm IR}$ is the
lensing-corrected infrared luminosity: ][]{scoville04}. For such
maximally-efficient star formation, a measurement of $L_{\rm IR}$
therefore corresponds to a determination of $M_{\rm SF}$. Using
canonical brightness temperature ratios for cold, quiescent ($r^{\rm
  (CQ)}_{(3-2)/(1-0)} = 0.3$) and warm, star-forming ($r^{\rm
  (SF)}_{(3-2)/(1-0)} = 1.0$) gas, their relative fractions in a
sub-thermally-excited reservoir can be computed as

\begin{center}
\begin{equation}\label{eq:mcqsf}
m_{\rm CQ/SF} = \frac{r^{\rm (SF)}_{(3-2)/(1-0)} - r_{(3-2)/(1-0)}}{r_{(3-2)/(1-0)} - r^{\rm (CQ)}_{(3-2)/(1-0)}}
\end{equation}
\end{center}

and the total molecular gas mass is calculated as

\begin{center}
\begin{equation}\label{eq:mgasI11}
 M_{\rm gas} = M_{\rm SF}(1+m_{\rm CQ/SF})
\end{equation}
\end{center}

Computing $M_{\rm gas}$ in this way provides an independent check on
the total gas mass, based on different assumptions from those that
underpin the X-factor method. The magnification-weighted mean value of
$M_{\rm gas, I11}$ for SMM\,J16359 is $(2.1 \pm 0.6) \times
10^9$\,M$_{\odot}$, and for SMM\,J02399 we detect $(41.0 \pm 14.4)
\times 10^9$\,M$_{\odot}$ of gas using this method. For SMM\,J14009 we
calculate $M_{\rm gas, I11} = (34.9 \pm 7.3) \times
10^9$\,M$_{\odot}$. All three values of $M_{\rm gas, I11}$ are $\sim
2\times$ lower than the equivalent values calculated via $X_{\rm CO}$,
which would imply $X_{\rm CO} \sim0.4$.  We remind the reader,
however, that $M_{\rm gas, I11}$ is calculated on the assumption that
SMGs have maximal SFE: if the SFE is \textit{less} than
500\,L$_{\odot}$\,M$_{\odot}^{-1}$, then a higher mass of star-forming
gas is required for a given $L_{\rm IR}$, raising the total gas mass
in the system accordingly.

We can then compare these gas masses to the dynamical masses, assuming
$M_{\rm dyn} = 2.1\,R\,\sigma_{\rm CO(1-0)}^{2}/G$
\citep{tacconi08}. Radii for our sources are derived by fitting a
two-dimensional Gaussian to each integrated \co\ \jonezero\ map. This
estimate assumes that the gas emission traces a virialised potential
well. Using the observed values of $\sigma_{\rm CO(1-0)}$ reported in
Table~2, we determine dynamical masses of $(18 \pm 4) \times
10^9$\,M$_{\odot}$, $(11 \pm 4) \times 10^9$\,M$_{\odot}$ and $(1560
\pm 550) \times 10^9$\,M$_{\odot}$ for SMM\,J14009, SMM\,J16359-B and
SMM\,J02399, implying firm upper-limits of $X_{\rm CO} < 0.3$, $X_{\rm
  CO} < 2.8$ and $X_{\rm CO} < 24$, respectively. The dynamical masses
derived from the two brighter images of SMM\,J16359 are
$\sim$3$\times$ higher than the best-estimate of the gas mass derived
via $X_{\rm CO} = 0.8$, consistent with the sample of
\citet{ivison11}. The extremely compact radius quoted for SMM\,J14009
is a result of the narrowness of the 2D Gaussian that best fits the
integrated map, itself a result of the source being only marginally
resolved on North-South baselines and completely unresolved on
East-West baselines. The gas in SMM\,J02399 was reported as having a
lensing-corrected extent of $\sim$25\,kpc in \citet{ivison10b} in
their C-configuration observations, corroborating the equally large
extent of $26 \pm 4$\,kpc derived from our D-configuration
observations, as reported in Table~2. The colossal value of $M_{\rm
  dyn}$ therefore likely implies that the assumption the gas traces a
virialised potential well is false.

\subsection{Gas excitation: observed brightness temperature ratios}

Having worked with canonical brightness temperature ratios to estimate
$M_{\rm gas}$, free from any assumptions about $X_{\rm CO}$, we now
compare our \co\ \jonezero\ spectra and the literature \jthreetwo\
spectra in order to determine the true excitation conditions of the
gas in our samples of galaxies.

The steepness of the observed radio continuum spectrum of SMM\,J02399
(\S\ref{subsect:radio-continuum}) is consistent with the claim that
this source harbours a powerful AGN. Such an AGN might be expected to
heat the surrounding gas via bombardment with X-rays
\citep{meijerink06} or through shock-heating as the powerful radio jet
drives into the surrounding molecular gas \citep{papadopoulos11,
  ivison12}, yet the $T_{\rm b}$ ratio detected for this source
$r_{(3-2)/(1-0)} = 0.58 \pm 0.14$ is entirely consistent with that of
less AGN-like SMGs. As \citet{ivison10b} observe, however, SMM\,J02399
appears to consist of three or more interacting galaxies embedded
within a larger reservoir of cold molecular gas. In particular, the
massive obscured starburst which gives rise to the extreme $L_{\rm
  IR}$ of this source is spatially offset from the
broad-absorption-line quasar.

Our improved sampling of the radio SED of SMM\,J14009 allows us to
determine that this source also has a steep radio continuum spectrum
(\S\ref{subsect:radio-continuum}), corroborating its classification as
an AGN. Additionally, SMM\,J14009 demonstrates a high $T_{\rm b}$
ratio, $r_{(3-2)/(1-0)} = 0.95 \pm 0.12$, which indicates the presence
of a warm gas phase given the excitation requirements of the \co\
\jthreetwo\ transition, $n_{\rm crit} \sim 10^4$\,cm$^{-3}$,
$E_{\rm u}/k_{\rm B} \sim 33$\,{\sc k}.

For SMM\,J16359, the magnification-weighted mean $T_{\rm b}$ ratio,
$r_{(3-2)/(1-0)} = 0.72 \pm 0.12$, is entirely consistent with that of
the SMG population \citep{harris10, ivison11}.

\begin{figure}
\includegraphics[width=9cm]{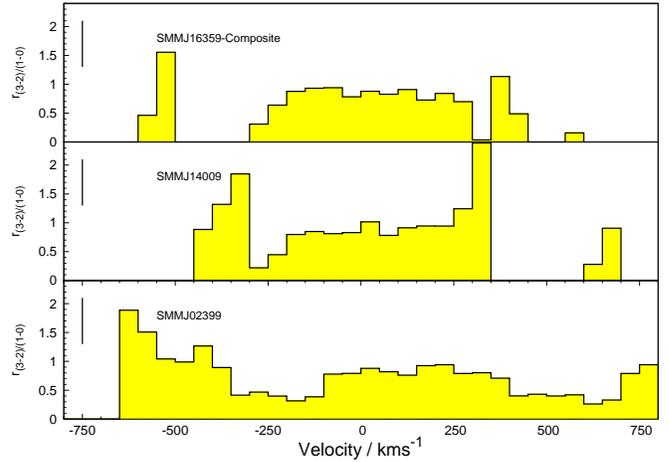}
\caption{\textit{Top -- bottom:} Plots of $T_{\rm b}$ ratio as a
  function of velocity for the composite SMM\,J16359 spectrum,
  SMM\,J14009 and SMM\,J02399. Due to poor signal/noise at the
  extremities of the line profiles, we only consider data within the
  central $\pm400$\,km\,s$^{-1}$ to be robust enough to draw any
  conclusions from. The composite spectrum for SMM\,J16359 is computed
  by taking the magnification-weighted mean of the individual spectra
  from images A, B and C of this triply imaged source. The
  observed velocity offset between \co\ \jonezero\ and \jthreetwo\
  spectra in SMM\,J02399 manifests itself as a super-thermal $T_{\rm
    b}$ ratios in the envelope of the velocity profile. Typical
  uncertainties are denoted by the error bar in the top-left of each
  plot.}
\label{fig:all-r3210}
\end{figure}

In addition to calculating source-averaged $T_{\rm b}$ ratios, our new
data give us the spatial and spectral resolution to assess
$r_{(3-2)/(1-0)}$ as functions of position and velocity. In
Fig.~\ref{fig:all-r3210} we illustrate the variation in gas excitation
across the observed line profiles for the three sources. The results
are only considered to be robust within the central
$\pm$400\,km\,s$^{-1}$ of each of the line profiles due to
deteriorating signal/noise at the edges of the line profiles;
nevertheless, in SMM\,J16359 we see clear structure in the plot of
$r_{(3-2)/(1-0)}$ versus velocity, with $r_{(3-2)/(1-0)} \sim 1$ for
regions in the centre of the velocity-space, rolling off to
sub-thermal values for the high-velocity wings of the line. These
variations in the velocity fields of each galaxy may point to spatial
variations in $r_{(3-2)/(1-0)}$, and are consistent with the observed
values of $\sigma_{(3-2)/(1-0)}$.

%
%

\begin{figure*}
\centerline{\psfig{file=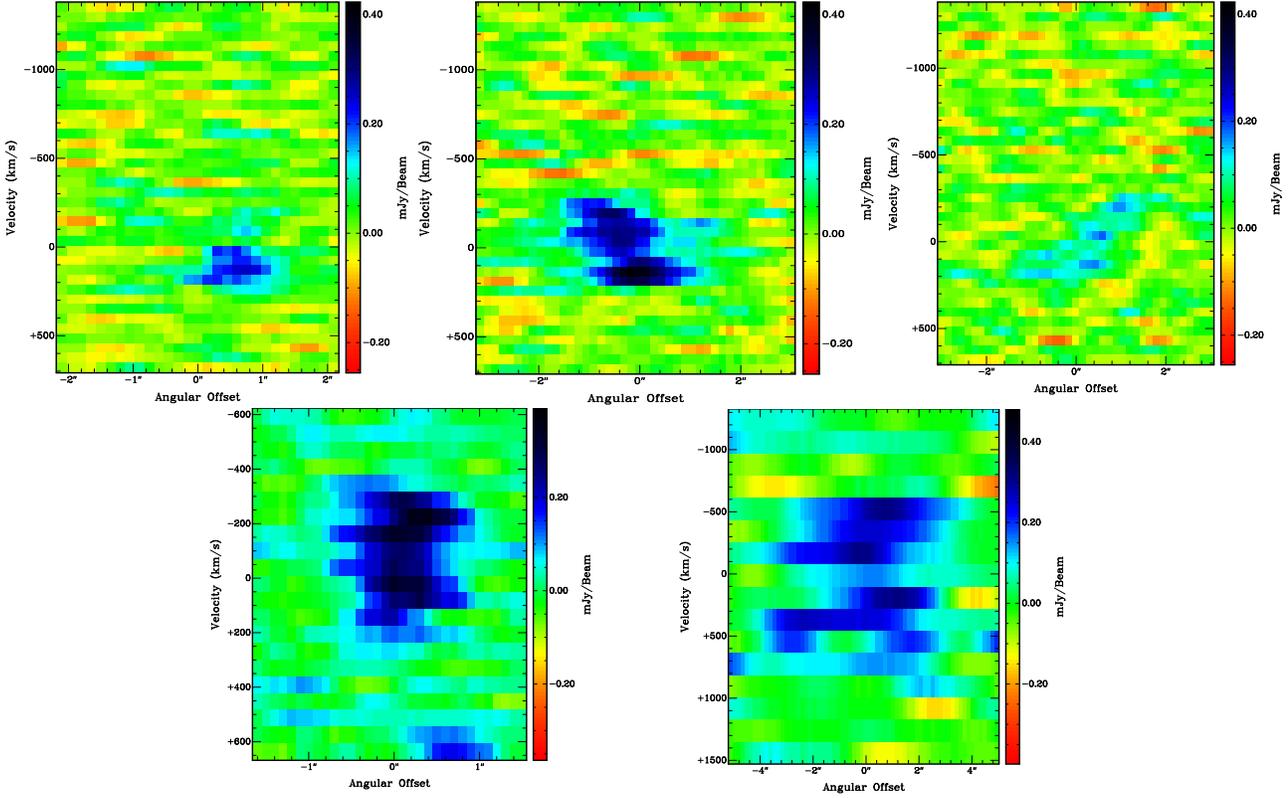,width=17cm}}
\caption{\textit{Clockwise from top left:} Position-velocity diagrams
  for SMM\,J16359-A, SMM\,J16359-B, SMM\,J16359-C, SMM\,J02399 and
  SMM\,J14009.}
\label{fig:PVDiagrams}
\end{figure*}

We attempt investigate the velocity structure of our galaxies by
constructing position-velocity (PV) diagrams for each source,
extracting a slice through the (right-ascension, declination,
velocity) data cube along the major axis of the galaxy. The resultant
PV diagrams are shown in Fig.~\ref{fig:PVDiagrams}, however their low
signal-to-noise prohibits us from modelling the emission.

%

\subsection{Infrared-derived star-formation rates}\label{sec:sfrir}

We present our infrared SEDs in Fig.~\ref{fig:seds} and use the
infrared luminosities reported in Table~3 with the truncation from
\citet{genzel03} of the \citet[hereafter K98]{kennicutt98} conversion
factor for a 1--100\,$\msol$ \citet{salpeter55} initial mass function
(IMF) to compute SFR$_{\rm IR}$:

\begin{center}
\begin{equation}\label{eq:sfr-from-ir}
\frac{{\rm SFR_{\rm IR}}}{1\,\msol \rm yr^{-1}} = \frac{L_{\rm
    IR}}{1.2 \times 10^{10}\,{\rm L}_{\odot}}
\end{equation}
\end{center}

In the case of SMM\,J02399, $L_{\rm IR}$ is thought to arise roughly
equally from the AGN and starburst components
\citep[e.g.][]{valiante07}. Adopting this AGN fraction for both
SMM\,J02399 and SMM\,J14009, Equation~\ref{eq:sfr-from-ir} yields
SFR$_{\rm IR} =340 \pm 40$ and $670 \pm 70\msol$\,yr$^{-1}$,
respectively, where the uncertainties are purely statistical (due to
$L_{\rm IR}$), and thus do not include the uncertainty in the K98
conversion factor.

\subsection{An independent probe of the SFR via optically-thin
  free-free emission}\label{sect:radio-continuum}

The SED longward of the FIR dust bump in galaxies is characterised by
three main components: (i) the Rayleigh-Jeans tail of the
infrared/sub-mm dust spectrum, (ii) thermalised free-free emission
arising from ionised H\,{\sc ii} regions, which traces the
massive/young stars ($> 5$\,$\msol$) capable of photoionising, and
(iii) synchrotron radiation arising from the electrons accelerated by
supernova remnants \citep[ hereafter C92]{condon92}. Sometimes (iv)
synchrotron radiation arising from plasma accelerated by an active AGN
may also contribute.

Our data and the new estimates of $L_{\rm IR}$ provide us with the
means to determine the SFR via two independent methods: (i) via
$L_{\rm IR}$ between 1--100\,$\msol$\,yr$^{-1}$ (K98) and (ii) via the
optically-thin free-free emission, probing SFR($\geq 5 \msol$) (C92).

We determine SFR$_{\rm radio}$ via Equation~\ref{eq:sfr-from-ff},
where $L_{\rm ff}$ is the free-free luminosity density at frequency
$\nu$, defined such that $L=4 \pi d^2 S_{\rm ff}$ for
lensing-corrected flux density, $S_{\rm ff}$. The free-free luminosity
is derived by collating the radio continuum data summarised in
Table~4, and fitting the relative contributions to the observed SED
arising from dust (an extrapolation of the Rayleigh-Jeans side of the
best-fitting CE01 infrared template), thermalised free-free emission
($S_{\rm ff} \propto \nu^{-0.1}$: C92) and synchrotron emission
($S_{\rm sync} \propto \nu^{- \alpha_{\rm sync}}$). The free
parameters are the flux-scaling of the synchrotron and free-free
emission, and also the spectral index, $\alpha_{\rm sync}$, of the
synchrotron component.

\begin{center}
\begin{equation}\label{eq:sfr-from-ff}
\biggl( \frac{L_{\rm ff}}{\rm W\,Hz^{-1}} \biggr) \sim 5.5 \times 10^{20}\biggl(\frac{\nu}{\rm GHz}\biggr)^{-0.1}\biggl[\frac{{\rm SFR}(M \geq 5 \msol)}{\msol \rm yr^{-1}}\biggr]
\end{equation}
\end{center}

%
\begin{table*}\label{tab:radio-observed}
  \centering
  \caption{Observed radio properties}
  \begin{tabular}{lccccccc}
    \hline
    \multicolumn{1}{l}{Target} &
    \multicolumn{1}{c}{$S_{\rm 1.4\,GHz}$} &
    \multicolumn{1}{c}{$S_{\rm 5\,GHz}$} &
    \multicolumn{1}{c}{$S_{\rm 8\,GHz}$} &
    \multicolumn{1}{c}{$S_{\rm 32\,GHz}$} &
    \multicolumn{1}{c}{$\alpha$} & \\
    \multicolumn{1}{l}{name} &
    \multicolumn{1}{c}{($\mu$Jy)} &
    \multicolumn{1}{c}{($\mu$Jy)} &
    \multicolumn{1}{c}{($\mu$Jy)} &
    \multicolumn{1}{c}{($\mu$Jy)} &
    \multicolumn{1}{c}{} & \\
     \hline
SMM\,J14009&529$^a$ $\pm$ 30&151 $\pm$ 17&91 $\pm$ 18&42 $\pm$ 20&0.96 $\pm$ 0.06\\
SMM\,J02399&526$^b$ $\pm$ 50&130 $\pm$ 22&N/A&57 $\pm$25&0.98 $\pm$ 0.14\\
    \hline
  \end{tabular}

{\small
Notes: $^a$\citet{ivison00}; $^b$\citet{ivison98}.}
\end{table*}

Using the 1.4-, 5- and 8-GHz data available for our two galaxies, we
measure spectral indices, $\alpha$ (where $S_{\nu} \propto
\nu^{\alpha}$), of $-0.96\pm 0.06$ and $-0.98\pm 0.14$ for SMM\,J14009
and SMM\,J02399, respectively, a little steeper than the mean value of
$-0.75 \pm 0.06$ measured by \citet{ibar10} between 610\,MHz and
1.4\,GHz for 44 radio-detected SMGs in the Lockman Hole, and steeper
too than the CE01 template that best fits each galaxy's
mid-IR-to-submm SED.

The contributions to $S_{\rm 32\,GHz}$ arising due to dust,
synchrotron and free-free emission are presented in Table~5. We
determine SFR($M \geq 5\,\msol$) = $290 \pm 85\,\msol$\,yr$^{-1}$ for
SMM\,J14009 and $200 \pm 20\,\msol$\,yr$^{-1}$ for SMM\,J02399, and
extrapolate down to 1\,$\msol$. Accounting for the unseen low-mass
stars using a \citet{salpeter55} IMF, we find 1--100$\msol$ SFR$_{\rm
  radio}$ of $630 \pm 190\,\msol$\,yr$^{-1}$ and $430 \pm
50\,\msol$\,yr$^{-1}$ for SMM\,J14009 and SMM\,J02399,
respectively. In common with the preceding section, these
uncertainties are statistical, and do not reflect the uncertainty in
the C92 conversion factor.

%
%
\begin{table*}\label{tab:radio-derived}
  \centering
  \caption{Radio source derived properties}
  \begin{tabular}{lcccccccc}
    \hline
    \multicolumn{1}{l}{Target} &
    \multicolumn{1}{c}{$S_{\rm Ka, sync}$} &
    \multicolumn{1}{c}{$\alpha_{\rm sync}$} &
    \multicolumn{1}{c}{$S_{\rm Ka, dust}$} &
    \multicolumn{1}{c}{$S_{\rm Ka, ff}$} &
    \multicolumn{1}{c}{SFR$_{\rm radio}^{a}$} &
    \multicolumn{1}{c}{SFR$_{\rm IR}^{a,b}$} & \\
    \multicolumn{1}{l}{name} &
    \multicolumn{1}{c}{($\mu$Jy)} &
    \multicolumn{1}{c}{} &
    \multicolumn{1}{c}{($\mu$Jy)} &
    \multicolumn{1}{c}{($\mu$Jy)} &
    \multicolumn{1}{c}{\sfr} &
    \multicolumn{1}{c}{\sfr} \\
     \hline
SMM\,J14009&22 $\pm$ 2&1.01 $\pm$ 0.09&8 $\pm$ 1&12 $\pm$ 5 &630 $\pm$ 190 & 670 $\pm$ 70\\
SMM\,J02399&16 $\pm$ 4&1.09 $\pm$ 0.13&18 $\pm$ 1&24 $\pm$ 6 & 430 $\pm$ 50 & 340 $\pm$ 40\\
    \hline
  \end{tabular}

  {\small
    Notes: $^a$SFRs calculated for 1-100\,$\msol$ Salpeter IMF;
             $^b$SFR$_{\rm IR}$ calculated on the assumption that the AGN and star formation contribute equally to $L_{\rm IR}$.}
\end{table*}

\subsection{The Lyman-break galaxy, Abell 2218 \#384}

Returning to the observations for SMM\,J16359, we searched for \co\
\jonezero\ in the LBG at \citep[R.A.\ 16:35:49.4, Dec.\ 66:13:07
J2000;][]{ebbels96} located behind the same cluster. We do not detect
this source, but set an upper limit using $S \leq
3\sigma_{n}\sqrt{\Delta v \Delta w}$ \citep{seaquist95}, where $\Delta
v$ is the velocity resolution of the spectrum, $\Delta w \sim
250$\,\kms\ is a typical LBG line width \citep{greve08} and the noise
in the primary beam-corrected VLA spectrum is $\sigma_{n}$. We
de-boost this by $\sim16\times$ to account for gravitational lensing
by the cluster \citep{ebbels96} and thus derive upper limits on the
\co\ \jonezero\ luminosity of $<
0.8\times10^{9}$\,Jy\,km\,s$^{-1}$\,pc$^{2}$ and on the gas mass of
$<6\times10^{8}\,\msol$, respectively, where we have assumed $X_{\rm
  CO} = 0.8$, supposedly appropriate for star-forming galaxies.

This gas mass limit is comparable to gas masses detected in two
similarly high-redshift LBGs by \citet{riechers10}, who determined
$M_{\rm gas} = (9.3\pm1.6)\times10^{8}\msol$ and $M_{\rm gas} =
(4.6\pm1.1)\times10^{8}\msol$\, for the same $X_{\rm CO}$, in the
Cosmic Eye (at $z=3.074$) and MS\,1512-cB58 (at $z=2.727$).

\section{Discussion and conclusions}

\subsection{The physical conditions of the molecular gas}

We have presented high-resolution VLA imaging of \co\ \jonezero\ in
three bright, lensed SMGs. Two of our sources harbour known AGN --
SMM\,J14009 and SMM\,J02399 -- and are brighter ($L_{\rm IR} \sim
10^{13}\,$L$_{\odot}$) than the unlensed sample of \citet{ivison11},
while the other source -- the triply imaged SMM\,J16359 -- is an order
of magnitude fainter ($L_{\rm IR}\sim6\times 10^{11}$\,L$_{\odot}$,
observed as being equally bright due to the high magnification
afforded by its lensing cluster, Abell 2218), and does not appear to
be dominated by AGN emission.

Comparing integrated maps of \co\ \jonezero\ and \jthreetwo\ emission,
we compute mean brightness temperature ratios of less than unity and
velocity width ratios $\sigma_{(3-2)/(1-0)} \gs 1$, both results
consistent with the findings of \citet{ivison11}: SMGs on average
experience sub-thermal excitation and exhibit broad \co\ \jonezero\
line widths.

We derive $L_{\rm IR}$ for our galaxies by collating literature data,
and then choosing from the SED library of \citet{chary01} the
best-fitting SED, which we integrate between rest-frame
8--1,000\,$\mu$m. Next, we compute molecular gas masses via two
methods: first by converting $L'_{\rm CO}$ to $M_{\rm gas}$ via an
X-factor of $X_{\rm CO} = M$(H$_{2}$)/$L'_{\rm CO(1-0)}$ and then, as
an independent check, using an approach that assumes SMGs form stars
with Eddington-limited SFE, which gives us a lower-limit to the
molecular gas mass. For each of our sources, the latter method yields
a gas mass $\sim2\times$ lower than that derived via $X_{\rm CO}$.

\subsection{Independent estimates of the SFR, and the role played by AGN}

We compare SFRs derived via two independent diagnostics: (i) SFR$_{\rm
  IR}$, as computed via $L_{\rm IR}$ (K98) and (ii) SFR$_{\rm radio}$,
computed from $L_{\rm ff}$, the optically-thin free-free emission
(C92) (Table~5).  Both of these diagnostics are unaffected by
interstellar extinction, however the former method is liable to
over-estimate the SFR if a significant fraction of $L_{\rm IR}$ arises
due to heating of dust by an AGN \citep{murphy11}. In contrast, the
decomposition of the radio SED into dust, synchrotron and free-free
contributions provides us with an SFR diagnostic that is directly
related to the photoionisation rate of young, massive stars,
relatively unaffected by AGN.

To derive SFR$_{\rm IR}$ in \S3.4, we assumed that $\sim$50 per cent
of $L_{\rm IR}$ was due to star formation, with the rest due to AGN
heating of the dust \citep[e.g.][]{frayer98, greve05, valiante07}.
Here, we use our new measurements of free-free emission -- which is
unaffected by the presence of an AGN -- to test this assumption. Using
Equation \ref{eq:sfr-from-ir} \textit{without} an AGN correction,
SMM\,J02399's SFR$_{\rm IR, upper}$ is $670\pm 70$\,\sfr. This is
$35\pm10$ per cent higher than SFR$_{\rm radio}$ for a 1--100-$\msol$
Salpeter IMF, implying a $\sim$35 per cent contribution to $L_{\rm
  IR}$ from its AGN. Similarly for SMM\,J14009, SFR$_{\rm IR, upper} =
1,340 \pm 140$\,\sfr, indicating that some $55\pm15$ per cent of
$L_{\rm IR}$ arises due to the AGN.

Recognising that the spatial extent traced by \co\ \jonezero\ for each
of our galaxies is almost certainly larger than the region in which
star formation is actively taking place, we can use our new
free-free-derived SFRs to place lower-limits on the global
star-formation rate surface density, $\Sigma_{\rm SFR} =
0.7\pm0.1\msol$\,yr$^{-1}$\,kpc$^{-2}$ for SMM\,J02399 and
$\Sigma_{\rm SFR} = 450\pm200\msol$\,yr$^{-1}$\,kpc$^{-2}$ for
SMM\,J14009, the former value again attributable to the colossal
extent of the \co\ \jonezero\ reservoir in SMM\,J02399. We note that
the peak $\Sigma_{\rm SFR}$ will be significantly higher than these
galaxy-averaged values.

At present, our measurement of SFR$_{\rm radio}$ -- and hence our
estimates of the AGN fraction and $\Sigma_{\rm SFR}$ -- are only as
good as the extrapolation of the best-fit SED to the IR/submm data,
via which we determine the dust contribution to the $\nu_{\rm obs}
\sim 115$-GHz flux density, itself a relatively low signal-to-noise
measurement. However, these uncertainties can be reduced significantly
via better sampling of the SED beyond the wavelengths probed by
\herschel. When complete, the VLA will offer 8\,GHz of simultaneous
bandwidth across 64 independently tunable sub-band pairs. Together
with the Atacama Large Millimeter Array, it will be possible to
measure the 350\,$\mu$m--20\,cm SEDs of these galaxies with an order
of magnitude better sensitivity than currently possible.

\section*{Acknowledgements}

We would like to express our immense gratitude to the VLA
commissioning team and all those that have helped to create this
remarkable facility. AT, IRS and RJI acknowledge support from
STFC. IRS also acknowledges support from a Leverhulme Senior
Fellowship. {\it Herschel} is an ESA space observatory with science
instruments provided by European-led Principal Investigator consortia
and with important participation from NASA. The National Radio Astronomy Observatory is a facility of the National Science Foundation operated under cooperative agreement by Associated Universities, Inc.

\FloatBarrier
\bibliographystyle{mnras} 
\bibliography{20120629}

\begin{thebibliography}{}

\bibitem[\protect\citeauthoryear{{Bothwell} et~al.}{{Bothwell}
  et~al.}{2012}]{bothwell12}
{Bothwell} M.~S. et~al., 2012, ArXiv e-prints

\bibitem[\protect\citeauthoryear{{Brown} \& {Vanden Bout}}{{Brown} \& {Vanden
  Bout}}{1991}]{brown91}
{Brown} R.~L.,  {Vanden Bout} P.~A., 1991, \aj, 102, 1956

\bibitem[\protect\citeauthoryear{{Chapman} et~al.}{{Chapman}
  et~al.}{2005}]{chapman05}
{Chapman} S.~C., {Blain} A.~W., {Smail} I.,  {Ivison} R.~J., 2005, \apj, 622,
  772

\bibitem[\protect\citeauthoryear{{Chary} \& {Elbaz}}{{Chary} \&
  {Elbaz}}{2001}]{chary01}
{Chary} R.,  {Elbaz} D., 2001, \apj, 556, 562

\bibitem[\protect\citeauthoryear{{Condon}}{{Condon}}{1992}]{condon92}
{Condon} J.~J., 1992, \araa, 30, 575

\bibitem[\protect\citeauthoryear{{Danielson} et~al.}{{Danielson}
  et~al.}{2011}]{danielson11}
{Danielson} A.~L.~R. et~al., 2011, \mnras, 410, 1687

\bibitem[\protect\citeauthoryear{{Downes} \& {Solomon}}{{Downes} \&
  {Solomon}}{1998}]{downes98}
{Downes} D.,  {Solomon} P.~M., 1998, \apj, 507, 615

\bibitem[\protect\citeauthoryear{{Ebbels} et~al.}{{Ebbels}
  et~al.}{1996}]{ebbels96}
{Ebbels} T.~M.~D., {Le Borgne} J.-F., {Pello} R., {Ellis} R.~S., {Kneib} J.-P.,
  {Smail} I.,  {Sanahuja} B., 1996, \mnras, 281, L75

\bibitem[\protect\citeauthoryear{{Finkelstein} et~al.}{{Finkelstein}
  et~al.}{2011}]{finkelstein11}
{Finkelstein} K.~D. et~al., 2011, \apj, 742, 108

\bibitem[\protect\citeauthoryear{{Frayer} et~al.}{{Frayer}
  et~al.}{1998}]{frayer98}
{Frayer} D.~T., {Ivison} R.~J., {Scoville} N.~Z., {Yun} M., {Evans} A.~S.,
  {Smail} I., {Blain} A.~W.,  {Kneib} J.-P., 1998, \apjl, 506, L7

\bibitem[\protect\citeauthoryear{{Genzel} et~al.}{{Genzel}
  et~al.}{2003}]{genzel03}
{Genzel} R., {Baker} A.~J., {Tacconi} L.~J., {Lutz} D., {Cox} P., {Guilloteau}
  S.,  {Omont} A., 2003, \apj, 584, 633

\bibitem[\protect\citeauthoryear{{Graham} \& {Liu}}{{Graham} \&
  {Liu}}{1995}]{graham95}
{Graham} J.~R.,  {Liu} M.~C., 1995, \apjl, 449, L29

\bibitem[\protect\citeauthoryear{{Greve} et~al.}{{Greve}
  et~al.}{2005}]{greve05}
{Greve} T.~R. et~al., 2005, \mnras, 359, 1165

\bibitem[\protect\citeauthoryear{{Greve} \& {Sommer-Larsen}}{{Greve} \&
  {Sommer-Larsen}}{2008}]{greve08}
{Greve} T.~R.,  {Sommer-Larsen} J., 2008, \aap, 480, 335

\bibitem[\protect\citeauthoryear{{Hainline} et~al.}{{Hainline}
  et~al.}{2011}]{hainline11}
{Hainline} L.~J., {Blain} A.~W., {Smail} I., {Alexander} D.~M., {Armus} L.,
  {Chapman} S.~C.,  {Ivison} R.~J., 2011, \apj, 740, 96

\bibitem[\protect\citeauthoryear{{Harris} et~al.}{{Harris}
  et~al.}{2010}]{harris10}
{Harris} A.~I., {Baker} A.~J., {Zonak} S.~G., {Sharon} C.~E., {Genzel} R.,
  {Rauch} K., {Watts} G.,  {Creager} R., 2010, \apj, 723, 1139

\bibitem[\protect\citeauthoryear{{Hempel} et~al.}{{Hempel}
  et~al.}{2008}]{hempel08}
{Hempel} A., {Schaerer} D., {Egami} E., {Pell{\'o}} R., {Wise} M., {Richard}
  J., {Le Borgne} J.-F.,  {Kneib} J.-P., 2008, \aap, 477, 55

\bibitem[\protect\citeauthoryear{{Ibar} et~al.}{{Ibar} et~al.}{2010}]{ibar10}
{Ibar} E., {Ivison} R.~J., {Best} P.~N., {Coppin} K., {Pope} A., {Smail} I.,
  {Dunlop} J.~S., 2010, \mnras, 401, L53

\bibitem[\protect\citeauthoryear{{Ivison} et~al.}{{Ivison}
  et~al.}{2011}]{ivison11}
{Ivison} R.~J., {Papadopoulos} P.~P., {Smail} I., {Greve} T.~R., {Thomson}
  A.~P., {Xilouris} E.~M.,  {Chapman} S.~C., 2011, \mnras, 412, 1913

\bibitem[\protect\citeauthoryear{{Ivison} et~al.}{{Ivison}
  et~al.}{2012}]{ivison12}
{Ivison} R.~J. et~al., 2012, ArXiv e-prints

\bibitem[\protect\citeauthoryear{{Ivison} et~al.}{{Ivison}
  et~al.}{2000}]{ivison00}
{Ivison} R.~J., {Smail} I., {Barger} A.~J., {Kneib} J.-P., {Blain} A.~W.,
  {Owen} F.~N., {Kerr} T.~H.,  {Cowie} L.~L., 2000, \mnras, 315, 209

\bibitem[\protect\citeauthoryear{{Ivison} et~al.}{{Ivison}
  et~al.}{1998}]{ivison98}
{Ivison} R.~J., {Smail} I., {Le Borgne} J.-F., {Blain} A.~W., {Kneib} J.-P.,
  {Bezecourt} J., {Kerr} T.~H.,  {Davies} J.~K., 1998, \mnras, 298, 583

\bibitem[\protect\citeauthoryear{{Ivison} et~al.}{{Ivison}
  et~al.}{2010}]{ivison10b}
{Ivison} R.~J., {Smail} I., {Papadopoulos} P.~P., {Wold} I., {Richard} J.,
  {Swinbank} A.~M., {Kneib} J.-P.,  {Owen} F.~N., 2010, \mnras, 404, 198

\bibitem[\protect\citeauthoryear{{Kennicutt}}{{Kennicutt}}{1998}]{kennicutt98}
{Kennicutt} R.~C., Jr., 1998, \apj, 498, 541

\bibitem[\protect\citeauthoryear{{Kneib}}{{Kneib}}{2010}]{kneib10}
{Kneib} J.-P., 2010, in {Macchetto} F.~D., ed, The Impact of HST on European
  Astronomy.
\newblock Springer, p. 183

\bibitem[\protect\citeauthoryear{{Kneib} et~al.}{{Kneib}
  et~al.}{2005}]{kneib05}
{Kneib} J.-P., {Neri} R., {Smail} I., {Blain} A., {Sheth} K., {van der Werf}
  P.,  {Knudsen} K.~K., 2005, \aap, 434, 819

\bibitem[\protect\citeauthoryear{{Kneib} et~al.}{{Kneib}
  et~al.}{2004}]{kneib04}
{Kneib} J.-P., {van der Werf} P.~P., {Kraiberg Knudsen} K., {Smail} I., {Blain}
  A., {Frayer} D., {Barnard} V.,  {Ivison} R., 2004, \mnras, 349, 1211

\bibitem[\protect\citeauthoryear{{Lutz} et~al.}{{Lutz} et~al.}{2005}]{Lutz05}
{Lutz} D., {Valiante} E., {Sturm} E., {Genzel} R., {Tacconi} L.~J., {Lehnert}
  M.~D., {Sternberg} A.,  {Baker} A.~J., 2005, \apjl, 625, L83

\bibitem[\protect\citeauthoryear{{Meijerink}, {Spaans} \& {Israel}}{{Meijerink}
  et~al.}{2006}]{meijerink06}
{Meijerink} R., {Spaans} M.,  {Israel} F.~P., 2006, \apjl, 650, L103

\bibitem[\protect\citeauthoryear{{Morton} \& {Noreau}}{{Morton} \&
  {Noreau}}{1994}]{morton94}
{Morton} D.~C.,  {Noreau} L., 1994, \apjs, 95, 301

\bibitem[\protect\citeauthoryear{{Murphy} et~al.}{{Murphy}
  et~al.}{2011}]{murphy11}
{Murphy} E.~J. et~al., 2011, \apj, 737, 67

\bibitem[\protect\citeauthoryear{{Negrello} et~al.}{{Negrello}
  et~al.}{2010}]{negrello10}
{Negrello} M. et~al., 2010, Science, 330, 800

\bibitem[\protect\citeauthoryear{{Papadopoulos} \& {Seaquist}}{{Papadopoulos}
  \& {Seaquist}}{1999}]{papadopoulos99}
{Papadopoulos} P.~P.,  {Seaquist} E.~R., 1999, \apj, 516, 114

\bibitem[\protect\citeauthoryear{{Papadopoulos} et~al.}{{Papadopoulos}
  et~al.}{2011}]{papadopoulos11}
{Papadopoulos} P.~P., {van der Werf} P., {Xilouris} E.~M., {Isaak} K.~G., {Gao}
  Y.,  {Muehle} S., 2011, ArXiv e-prints

\bibitem[\protect\citeauthoryear{{Pilbratt} et~al.}{{Pilbratt}
  et~al.}{2010}]{pilbratt10}
{Pilbratt} G.~L. et~al., 2010, \aap, 518, L1

\bibitem[\protect\citeauthoryear{{Riechers} et~al.}{{Riechers}
  et~al.}{2010}]{riechers10}
{Riechers} D.~A., {Carilli} C.~L., {Walter} F.,  {Momjian} E., 2010, \apjl,
  724, L153

\bibitem[\protect\citeauthoryear{{Rowan-Robinson} et~al.}{{Rowan-Robinson}
  et~al.}{1991}]{rowan-robinson91}
{Rowan-Robinson} M. et~al., 1991, \nat, 351, 719

\bibitem[\protect\citeauthoryear{{Salpeter}}{{Salpeter}}{1955}]{salpeter55}
{Salpeter} E.~E., 1955, \apj, 121, 161

\bibitem[\protect\citeauthoryear{{Scoville}}{{Scoville}}{2004}]{scoville04}
{Scoville} N., 2004, in Astronomical Society of the Pacific Conference Series,
  Vol. 320, {Aalto} S., {Huttemeister} S.,  {Pedlar} A., eds, The Neutral ISM
  in Starburst Galaxies, p. 253

\bibitem[\protect\citeauthoryear{{Seaquist}, {Ivison} \& {Hall}}{{Seaquist}
  et~al.}{1995}]{seaquist95}
{Seaquist} E.~R., {Ivison} R.~J.,  {Hall} P.~J., 1995, \mnras, 276, 867

\bibitem[\protect\citeauthoryear{{Smail}, {Ivison} \& {Blain}}{{Smail}
  et~al.}{1997}]{smail97}
{Smail} I., {Ivison} R.~J.,  {Blain} A.~W., 1997, \apjl, 490, L5

\bibitem[\protect\citeauthoryear{{Smail} et~al.}{{Smail}
  et~al.}{2002}]{smail02}
{Smail} I., {Ivison} R.~J., {Blain} A.~W.,  {Kneib} J.-P., 2002, \mnras, 331,
  495

\bibitem[\protect\citeauthoryear{{Solomon}, {Radford} \& {Downes}}{{Solomon}
  et~al.}{1992}]{solomon92}
{Solomon} P.~M., {Radford} S.~J.~E.,  {Downes} D., 1992, \nat, 356, 318

\bibitem[\protect\citeauthoryear{{Tacconi} et~al.}{{Tacconi}
  et~al.}{2008}]{tacconi08}
{Tacconi} L.~J. et~al., 2008, \apj, 680, 246

\bibitem[\protect\citeauthoryear{{Valiante} et~al.}{{Valiante}
  et~al.}{2007}]{valiante07}
{Valiante} E., {Lutz} D., {Sturm} E., {Genzel} R., {Tacconi} L.~J., {Lehnert}
  M.~D.,  {Baker} A.~J., 2007, \apj, 660, 1060

\bibitem[\protect\citeauthoryear{{Wei{\ss}} et~al.}{{Wei{\ss}}
  et~al.}{2005}]{weiss05}
{Wei{\ss}} A., {Downes} D., {Walter} F.,  {Henkel} C., 2005, \aap, 440, L45

\bibitem[\protect\citeauthoryear{{Wei{\ss}} et~al.}{{Wei{\ss}}
  et~al.}{2009}]{weiss09}
{Wei{\ss}} A., {Ivison} R.~J., {Downes} D., {Walter} F., {Cirasuolo} M.,
  {Menten} K.~M., 2009, \apjl, 705, L45

\end{thebibliography}
\bsp

\label{lastpage}

\end{document}